\documentclass[useAMS,usenatbib]{mn2e}
\usepackage{amssymb}
\usepackage{amsmath}
\usepackage{graphicx}

\title[]{Asteroseismology of the DOV star PG 1159-035}
\author[Y. H. Chen]{Y. H. Chen$^{1,2,3}$\thanks{E-mail: yanhuichen1987@126.com}\\
$^{1}$Institute of Astrophysics, Chuxiong Normal University, Chuxiong 675000, China\\
$^{2}$School of Physics and Electronical Science, Chuxiong Normal University, Chuxiong 675000,China\\
$^{3}$Key Laboratory for the Structure and Evolution of Celestial Objects, Chinese Academy of Sciences, P.O. Box 110, Kunming 650011, China\\}

\begin{document}

\date{Accepted: }

\pagerange{\pageref{firstpage}--\pageref{lastpage}} \pubyear{????}

\maketitle

\label{firstpage}

\begin{abstract}

Grids of DOV star models are evolved by \texttt{WDEC} with fixed atmospheric constituent to the spectral values of $X_{C}/X_{He}/X_{O}$ = 50/33/17. The core compositions are from white dwarf models at highest $T_{eff}$ evolved by \texttt{MESA}. The eigenfrequencies are calculated and used to fit the observed modes. Based on 264.1 hours of photometric observations on PG 1159-035, Winget et al. identified 125 individual frequencies. Costa et al. identified 198 pulsation modes for PG 1159-035 according to the WET photometric data from 1983, 1985, 1989, and 2002. Both of them derived frequency splitting values of $\delta\sigma_{l=1}$ $\sim$ 4.2\,$\mu$Hz and $\delta\sigma_{l=2}$ $\sim$ 6.9\,$\mu$Hz. According to the values of $\delta\sigma_{l=1}$ and $\delta\sigma_{l=2}$, 20 triplets and 9 quintuplets are selected and used to constrain the fitting models. Our optimal model has $T_{eff}$ = 129000\,K, $M_{*}$ = 0.63\,$M_{\odot}$, log$g$ = 7.59, log($M_{\rm env}/M_{*}$) = -5.0, and $\sigma_{RMS}$ = 1.97\,s. The values of $T_{eff}$ and log$g$ are consistent with that values of C\'orsico et al.. The calculated modes of minimum rate of period change correspond to modes with maximum kinetic energy distributed in the envelope. The observed rates of period change with positive and negative values can also be partially reproduced. In particular, there are negative rates of period change for the calculated modes from our optimal model, which is not found in previous work.

\end{abstract}


\section{Introduction}

The DOV stars, also known as GW Vir or pulsating PG 1159 stars, are hot hydrogen (H) deficient stars. Something has to have happened to PG 1159 stars during the planetary nebula phase to eject the H-rich layers, leaving behind a helium/carbon/oxygen (He/C/O) dominated atmosphere without exposing the C/O core. Werner \& Herwig (2006) reported that the PG 1159 stars probably experienced a late helium-shell thermal pulse leading to the H-deficient atmosphere. Seismology of PG 1159 stars offer clues on this process. The instability strip for DOV stars ranges from 200,000\,K to 75,000\,K. The logarithm of gravitational acceleration is from 5.5 to 8.0 in centimeter-gram-second system. DOV stars usually have many observed modes, and therefore are good candidates for asteroseismological study. In addition, the PG 1159 stars are probably progenitors of some DBV stars (Dehner \& Kawaler 1995). Therefore, the study on PG 1159 stars is of great significance to the study of the structure and evolution of stars.

In the Palomar Green Survey, a survey to search for ultraviolet excess objects, the star PG 1159-035 was identified as a faint blue object by Green (1977). Based on non-LTE line blanketed model atmospheres, Jahn et al. (2007) reported that the effective temperature ($T_{eff}$) was 140000$\pm$5000\,K for PG 1159-035. The gravitational acceleration (log$g$) was 7.0$\pm$0.5 for PG 1159-035 with stellar mass from 0.6\,$M_{\odot}$ to 0.7\,$M_{\odot}$ (Werner et al. 1991). Werner et al. (1991) also derived a main atmospheric constituent by mass fraction of C = 50\%, He = 33\%, and O = 17\%. The recent spectral study shows that PG 1159-035 has a detailed atmospheric constituent by mass fraction of C = 48\%, He = 32\%, O = 17\%, Ne = 2\%, H $\leq$ 2\%, N = 0.1\%, and few Fe (solar abundance), Si, P, S, F (Werner et al. 2011).

PG 1159-035 was also identified as a pulsating star by McGraw et al. (1979) according to the detection of two closely spaced periods. The star was observed several times in 1983, 1984, and 1985, and at least 8 eigen-modes were identified (Winget et al. 1985). For the Whole Earth Telescope (WET) runs in 1989 and 1993, PG 1159-035 was the primary target (Winget et al. 1991, Costa et al. 2003). For the WET runs in 1990, 2000, and 2002, PG 1159-035 was the secondary target (Bruvold 1993, Costa et al. 2003). Costa et al. (2008) reviewed all available WET photometric data (1983, 1985, 1989, 1993, and 2002) and identified a total of 198 pulsation modes for PG 1159-035. A rich frequency spectrum with numerous triplets and quintuplets were identified. For the work of model fitting, Kawaler \& Bradley (1994) made detailed fits to the 20 observed $l$ = 1 modes identified by Winget et al. (1991). They obtained a best fit model with $T_{eff} \approx 136,000$~K, $M_{\rm *}$ = 0.59$\pm$0.01\,$M_{\odot}$, log($M_{\rm env}/M_{*}$) = -2.40, and $X_{He}$ $\thickapprox$ 0.27 at the surface. By fitting the observed modes identified by Winget et al. (1991) and Costa et al. (2008), C\'orsico et al. (2008) derived their best fitting model with $T_{eff}$ $\approx$ 128000\,K, $M_{\rm *}$ $\approx$ 0.565\,$M_{\odot}$, log($M_{\rm env}/M_{*}$) = -1.52, and $X_{He}$ $\thickapprox$ 0.42 (C/He/O = 34/42/24, abundances by mass). We try to evolve theoretical DOV star models with fixed atmospheric constituent ($X_{C}/X_{He}/X_{O}$ = 50/33/17, Werner et al. 1991). We then make detailed period to period model fittings and perform our asteroseismological study on PG 1159-035.

We show the input physics and model calculations in Sect. 2. In Sect. 3, we perform an asteroseismology of PG 1159-035, including the mode selections, the model fittings, the mode trapping effect, and the rate of period change. At last, a discussion and conclusions are displayed in Sect. 4.

\section{Input Physics and Model Calculations}

The White Dwarf Evolution Code (\texttt{WDEC}) is a fast and versatile stellar evolution code designed for cooling white dwarf stars. \texttt{WDEC} does not have nuclear reactions, mass loss, and other previous stellar evolution processes. Depending on the starter model, it evolves hot white dwarf models from the pre-white dwarf phase to the effective temperature we need (Bischoff-Kim et al. 2018). The effective temperatures of initial input models ($\sim$100000\,K) are not hot enough for hot DOV stars. The \texttt{MESA} (Modules for Experiments in Stellar Astrophysics; Paxton et al. 2011) can evolve a star from the pre-main sequence stage to the white dwarf stage. In order to obtain evolved core compositions, the core compositions of white dwarfs evolved by \texttt{MESA} were taken out and added into \texttt{WDEC} to evolve white dwarf models (see Chen \& Li 2014 and their recent papers). We take the place of maximum carbon abundance ($X_{C}$ = 0.65 uniformly) as the boundary of the C/O core. The core composition contains the mass, radius, luminosity, pressure, temperature, entropy, and C abundance. The oxygen abundance is $X_{O}$ = 1.0 - $X_{C}$ in the core. We use the core structure of hot pre-white dwarfs evolved by \texttt{MESA} as starter models for \texttt{WDEC} and use them to evolve hot DOV models. The core mass is taken as the relative mass (log($M_{r}/M_{*}$)), so the core composition from white dwarfs evolved by \texttt{MESA} can be applied to white dwarfs of different masses evolved by \texttt{WDEC} in equal proportion.

\begin{table}
\begin{center}
\caption{Table of masses, including the main-sequence stars, white dwarfs evolved by \texttt{MESA} together with their highest effective temperatures and core masses (the boundary at $X_{c}$=0.65), and corresponding white dwarf masses evolved by \texttt{WDEC}.}
\begin{tabular}{llllll}
\hline
ID           &$MS$         &$WD(\texttt{MESA})$($T_{eff}$)           &$M_{core}(\texttt{MESA})$       &$WD(\texttt{WDEC})$     \\
\hline
             &($M_{\odot}$)&($M_{\odot}$)                            &($M_{\odot}$)                   &($M_{\odot}$)           \\
\hline
1            &3.0          &0.579 ($\sim$139100\,K)                  &0.548                      &0.57-0.65               \\
2            &3.2          &0.599 ($\sim$146400\,K)                  &0.571                      &0.57-0.65               \\
3            &3.4          &0.627 ($\sim$148300\,K)                  &0.604                      &0.57-0.65               \\
4            &3.5          &0.652 ($\sim$171900\,K)                  &0.629                      &0.57-0.65               \\
5            &3.6          &0.675 ($\sim$174600\,K)                  &0.655                      &0.57-0.65               \\
\hline
\end{tabular}
\end{center}
\end{table}

\begin{figure}
\begin{center}
\includegraphics[width=9.0cm,angle=0]{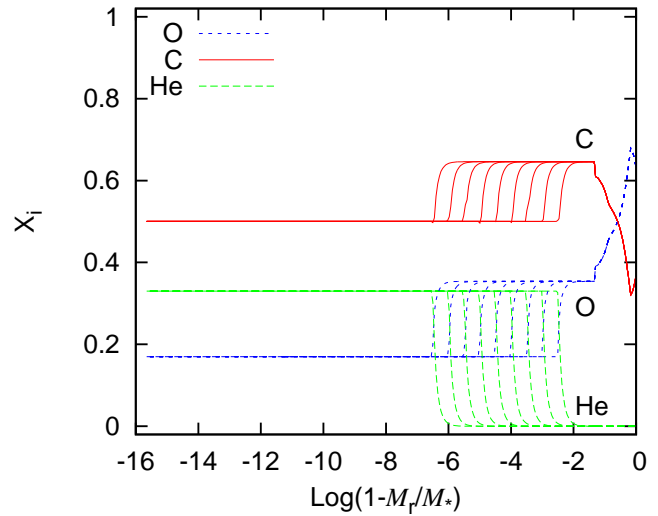}
\end{center}
\caption{Diagram of composition abundances of models with $T_{eff}$ = 129000\,K, $M_{\rm *}$ = 0.63\,$M_{\odot}$, and log($M_{env}/M_{*}$) = -2.5 to -6.5}
\end{figure}

We downloaded and installed the 6208 version of \texttt{MESA}. In a module '$make\_co\_wd$', a grid of main-sequence (MS) stars are evolved to be white dwarfs, see Table 1. For example, a 3.0\,$M_{\odot}$ MS star in \texttt{MESA} evolves to become a 0.579\,$M_{\odot}$ white dwarf with a maximum temperature $T_{eff}$ of $\sim$139100\,K. The white dwarf core mass is 0.548\,$M_{\odot}$ and has a core carbon abundance $X_{C}$ of 0.65. The core is taken from  \texttt{MESA} and used as a starter model for \texttt{WDEC}. We define the stellar evolution of 3.0\,$M_{\odot}$, 3.2\,$M_{\odot}$, 3.4\,$M_{\odot}$, 3.5\,$M_{\odot}$, and 3.6\,$M_{\odot}$ as ID 1, 2, 3, 4, and 5 respectively.

\texttt{WDEC} calculates the white dwarf cooling process. The equation of state (EOS) are from Lamb (1974), Fontaine et al. (1977), and Saumon et al. (1995). It has EOS tables of nearly pure H, pure He, and pure C. There is no EOS table of pure O. The DOV star PG 1159-035 has an atmospheric constituent of $X_{C}/X_{He}/X_{O}$ = 50/33/17 by mass, so we need an EOS table of pure O. In the file path 'mesa/data/eosPT\_data', there is a file named 'mesa-eosPT\_100z00x\_CO\_0.data'. The file contains EOS data of 50\% C and 50\% O. By linearly calculating the EOS data of 50\% C and 50\% O, and the EOS data of pure C, we create an EOS table of pure O. The EOS tables of pure He, pure C, and pure O in \texttt{WDEC} are used to evolve DOV star models. \texttt{WDEC} uses the OPAL radiative opacities, while the conductive opacities are from Itoh et al. (1983) and Hubbard \& Lampe (1969). An updated OPAL Rosseland mean opacity table of pure O (Iglesias \& Rogers 1996) was added into \texttt{WDEC}. We evolve DOV models with the spectroscopic abundance of $X_{C}/X_{He}/X_{O}$ = 50/33/17, and the portion of the stellar model with this composition is designated as the envelope mass $M_{env}$. The Ledoux term of the Brunt-Vi\"as\"al\"a frequency is computed taking all three elements into account. We calculate the opacity and EOS for the C/He and C/O mixtures separately and combine them according to their envelope abundances. The stellar mass is set as $M_{*}$. We evolve models with log($M_{env}/M_{*}$) ranging from -2.5 to -6.5 in steps of 0.5 as shown in Fig. 1. A grid of DOV models was computed with masses between $0.57$ to $0.65 M_{\odot}$ in steps of $0.01 M_{\odot}$, as shown in Table 1 (the last column).

\begin{table}
\begin{center}
\caption{The white dwarf masses with highest $T_{eff}$ values evolved by \texttt{WDEC}. They are from the evolving scenario of ID 3 in Table 1. Log($M_{env}/M_{*}$) is -5.0 for those models.}
\begin{tabular}{llllll}
\hline
ID           &$MS$         &$WD(\texttt{MESA})$           &$M_{core}(\texttt{MESA})$       &$WD(\texttt{WDEC})$($T_{eff}$)  \\
\hline
             &($M_{\odot}$)&($M_{\odot}$)                 &($M_{\odot}$)                   &($M_{\odot}$)                   \\
\hline
3            &3.4          &0.627                            &0.604                      &0.57(121600\,K)               \\
3            &3.4          &0.627                            &0.604                      &0.58(124400\,K)               \\
3            &3.4          &0.627                            &0.604                      &0.59(127200\,K)               \\
3            &3.4          &0.627                            &0.604                      &0.60(130000\,K)               \\
3            &3.4          &0.627                            &0.604                      &0.61(132800\,K)               \\
3            &3.4          &0.627                            &0.604                      &0.62(135600\,K)               \\
3            &3.4          &0.627                            &0.604                      &0.63(138600\,K)               \\
3            &3.4          &0.627                            &0.604                      &0.64(141400\,K)               \\
3            &3.4          &0.627                            &0.604                      &0.65(144400\,K)               \\
\hline
\end{tabular}
\end{center}
\end{table}

For each evolving scenario in Table 1, the core of a white dwarf evolved by \texttt{MESA} is used to proportionally evolve white dwarfs of $M_{\rm *}$ = 0.57-0.65\,$M_{\odot}$. The larger the white dwarf mass evolved by \texttt{WDEC}, the higher the effective temperature of the white dwarfs that can be evolved. In Table 2, a white dwarf of $M_{\rm *}$ = 0.57\,$M_{\odot}$ has a highest $T_{eff}$ of 121600\,K, while a white dwarf of $M_{\rm *}$ = 0.65\,$M_{\odot}$ has a highest $T_{eff}$ of 144,400\,K. The hydrogen abundance is set as infinitely small with log($M_{\rm H}/M_{*}$) = -200. Then, grids of DOV star models are evolved. We numerically solve the full equations of linear and adiabatic oscillation on those DOV star models based on the pulsation code of Li (1992a, 1992b). The eigenfrequencies can be scanned for the models. They are used to fit the observed modes for PG 1159-035.

\section{Asteroseismology of PG 1159-035}

The calculated modes are used to fit the observed modes for PG 1159-035. In subsection 3.1, we make mode selections for the observed modes identified by Winget et al. (1991) and Costa et al. (2008) according to the frequency splitting values. In subsection 3.2, we perform detailed period to period model fittings on the observed modes. We studied the mode trapping effect and the rate of period change in subsection 3.3.

\subsection{The mode selections for PG 1159-035}

PG 1159-035 had been observed many times since 1979. With 264.1 hours of photometric observations in 1989, Winget et al. (1991) made detailed mode identifications on PG 1159-035 (Table 3 of their paper). They derived an average frequency splitting value of 4.22\,$\mu$Hz for $l$ = 1 modes ($\delta\sigma_{l=1}$) and 6.92\,$\mu$Hz for $l$ = 2 modes ($\delta\sigma_{l=2}$). The ratio of 4.22/6.92=0.61 agrees well with the theoretical calculations of $\delta\sigma_{l=1}$/$\delta\sigma_{l=2}$=0.60. The observed average period spacing is $\Delta P_{l=1}$=21.5\,s for $l$ = 1 modes, and $\Delta P_{l=2}$=12.5\,s for $l$ = 2 modes. The ratio of 21.5/12.5=1.72 agrees well with the theoretical calculations of $\Delta P_{l=1}$/$\Delta P_{l=2}$=$\sqrt{3}$=1.73. Costa et al. (2008) used the available WET photometric data from 1983, 1989, 1993, and 2002 to identify 198 pulsation modes for PG 1159-035 (Table 4 and Table 12 of their paper). They also derived $\delta\sigma_{l=1}$=4.13\,$\mu$Hz, $\delta\sigma_{l=2}$=6.90\,$\mu$Hz, $\Delta P_{l=1}$=21.43\,s, and $\Delta P_{l=2}$=12.38\,s. According to the mode identifications of Winget et al. (1991) and Costa et al. (2008), we list the modes of three components for $l$ = 1 modes in Table 3, and five or four components for $l$ = 2 modes in Table 4.

There are 41 overtones in Table 3 and 4. We showed the frequency splitting values among components for each mode. According to the frequency splitting values for $l$ = 1 modes ($\delta\sigma_{l=1}$ $\sim$ 4.2\,$\mu$Hz), 20 modes were selected in Table 3. The value of $\delta\sigma_{l=1}$ is very large for $f_{1}$ and $f_{23}$, and very small for $f_{24}$ and $f_{25}$. We did not use them to constrain the models. For the mode of $f_{9}$, the value of $\delta\sigma_{l=1}$ was very large identified by Costa et al. (2008). For example, in Table 3 of Winget et al. (1991), there are independent frequencies of 1716.37, 1719.08, 1720.30, 1724.24, 1726.59\,$\mu$Hz. In Table 4 and 12 of Costa et al. (2008), there are independent frequencies of 1708.64, 1718.18, 1723.13, 1726.76, 1726.79, 1736.02, 1736.05\,$\mu$Hz. Therefore, we did not select the $l$ = 1 mode $f_{9}$ to constrain the models. The values of $\delta\sigma_{l=1}$ are near 4.22\,$\mu$Hz for the modes identified by Winget et al. (1991). For the modes identified by Costa et al. (2008), the dispersion of $\delta\sigma_{l=1}$ is large. We select the modes of $f_{2}$-$f_{21}$ (except $f_{9}$) from Winget et al. (1991), and $f_{22}$ from Costa et al. (2008) to constrain the models.

According to the frequency splitting values for $l$ = 2 modes ($\delta\sigma_{l=2}$ $\sim$ 6.9\,$\mu$Hz), 9 overtones were selected in Table 4. There is a frequency splitting value of 10.60\,$\mu$Hz for $f_{30}$, 3.18\,$\mu$Hz (11.69\,$\mu$Hz) for $f_{38}$, 9.58\,$\mu$Hz (4.71\,$\mu$Hz) for $f_{39}$, 9.26\,$\mu$Hz for $f_{40}$, and 18.24\,$\mu$Hz for $f_{41}$. In addition, there are four components for these modes. We did not select these $l$ = 2 modes to constrain the models. There is a frequency splitting value of 9.58\,$\mu$Hz for $f_{35}$. The component of 1937.87\,$\mu$Hz for $f_{35}$ was identified as an $l$ = 1 mode by Winget et al. (1991). Because of the uncertain $l$, we did not select $f_{35}$ to constrain the models. There is a frequency splitting value of 8.91\,$\mu$Hz for $f_{36}$. In Table 3 of Winget et al. (1991), there are independent frequencies of 1732.59, 1738.74, 1747.65, 1749.68\,$\mu$Hz. In Table 12 of Costa et al. (2008), there are independent frequencies of 1743.10 and 1750.73\,$\mu$Hz. Together with the mode identifications for $f_{9}$ in Table 3, we did not select the modes of $f_{9}$ and $f_{36}$ to constrain the models. The values of $\delta\sigma_{l=2}$ are basically around 6.92\,$\mu$Hz for the mode identifications of Winget et al. (1991). We select the modes of $f_{26}$-$f_{37}$ (except $f_{30}$, $f_{35}$, and $f_{36}$) from Winget et al. (1991) to constrain the models.

A reliable mode identification can effectively constrain the theoretical models. However, the detailed mode identifications are very complex, especially when the adjacent observed modes have large frequency interval dispersion. According to the frequency splitting values, we select 20 $l$ = 1 modes and 9 $l$ = 2 modes to constrain the models. It is a preliminary work to study PG 1159-035.

\begin{table*}
\begin{center}
\caption{The mode selections from Winget et al. (1991) and Costa et al. (2008) for $l$ = 1}
\begin{tabular}{lccccccccccccccc}
\hline
         &\multicolumn{3}{c}{Winget et al. (1991)}&\multicolumn{3}{c}{Costa et al. (2008)}   & Selected Period &         &\multicolumn{3}{c}{Costa et al. (2008)}   & Selected Period \\
\hline
ID       & Freq.    &$\delta\sigma$  & Per.       & Freq.    &$\delta\sigma$ & Per.          &$P_{\rm obs}$    &ID       & Freq.    &$\delta\sigma$ & Per.          &$P_{\rm obs}$    \\
$l$ = 1  &($\mu$Hz) &($\mu$Hz)       &(s)         &($\mu$Hz) &($\mu$Hz)      &(s)            &(s)              &         &($\mu$Hz) &($\mu$Hz)      &(s)            &(s)              \\
\hline
         &          &                &            &2565.94   &6.19           &389.72         &                 &         &1166.36   &5.89           &857.37         &                 \\
$f_{1}$  &          &                &            &2562.13   &               &390.30         &                 &$f_{22}$ &1160.47   &               &861.72         &861.72           \\
         &          &                &            &2558.59   &6.46           &390.84         &                 &         &1155.96   &4.51           &865.08         &                 \\
         &2330.24   &4.89            &429.14      &2323.53   &10.70          &430.38         &                 &         &1139.38   &7.73           &877.67         &                 \\
$f_{2}$  &2325.35   &                &430.04      &2312.83   &               &390.30         &430.04           &$f_{23}$ &1131.65   &               &883.67         &                 \\
         &2321.04   &4.31            &430.84      &2303.35   &9.48           &390.84         &                 &         &1124.02   &8.63           &889.66         &                 \\
         &2223.20   &4.46            &449.80      &2218.13   &6.03           &450.83         &                 &         &1083.20   &2.48           &923.19         &                 \\
$f_{3}$  &2218.74   &                &450.71      &2212.10   &               &452.06         &450.71           &$f_{24}$ &1080.72   &               &925.31         &                 \\
         &2214.38   &4.36            &451.59      &2206.34   &5.76           &453.24         &                 &         &1078.07   &2.65           &927.58         &                 \\
         &2133.15   &3.53            &468.79      &          &               &               &                 &         &1060.43   &2.24           &943.01         &                 \\
$f_{4}$  &2129.62   &                &469.57      &2118.29   &               &472.08         &469.57           &$f_{25}$ &1058.19   &               &945.01         &                 \\
         &2124.24   &5.38            &470.76      &2103.27   &15.02          &475.45         &                 &         &1055.51   &2.68           &947.41         &                 \\
         &2025.13   &4.33            &493.80      &2025.15   &4.34           &493.79         &                 &         &          &               &               &                 \\
$f_{5}$  &2020.80   &                &494.85      &2020.81   &               &494.85         &494.85           &         &          &               &               &                 \\
         &2016.46   &4.34            &495.92      &2016.13   &4.68           &496.00         &                 &         &          &               &               &                 \\
         &1937.83   &4.28            &516.04      &1937.83   &4.19           &516.04         &                 &         &          &               &               &                 \\
$f_{6}$  &1933.55   &                &517.18      &1933.64   &               &517.16         &517.18           &         &          &               &               &                 \\
         &1929.32   &4.23            &518.32      &1929.42   &4.22           &518.29         &                 &         &          &               &               &                 \\
         &1862.58   &4.38            &536.89      &1862.47   &4.22           &536.92         &                 &         &          &               &               &                 \\
$f_{7}$  &1858.20   &                &538.16      &1858.25   &               &538.14         &538.16           &         &          &               &               &                 \\
         &1854.04   &4.16            &539.36      &1854.12   &4.13           &539.34         &                 &         &          &               &               &                 \\
         &1794.88   &4.18            &557.14      &1794.91   &3.24           &557.13         &                 &         &          &               &               &                 \\
$f_{8}$  &1790.70   &                &558.44      &1791.67   &               &558.14         &558.44           &         &          &               &               &                 \\
         &1786.24   &4.46            &559.84      &1786.64   &5.03           &559.71         &                 &         &          &               &               &                 \\
         &1724.24   &3.94            &579.97      &1736.02   &9.26           &576.03         &                 &         &          &               &               &                 \\
$f_{9}$  &1720.30   &                &581.29      &1726.76   &               &579.12         &                 &         &          &               &               &                 \\
         &1716.37   &3.93            &582.62      &1718.18   &8.58           &581.67         &                 &         &          &               &               &                 \\
         &1662.66   &4.41            &601.44      &1662.66   &4.41           &601.44         &                 &         &          &               &               &                 \\
$f_{10}$ &1658.25   &                &603.04      &1658.25   &               &603.04         &603.04           &         &          &               &               &                 \\
         &1653.66   &4.59            &604.72      &1653.66   &4.59           &604.72         &                 &         &          &               &               &                 \\
         &1609.07   &2.90            &621.48      &1609.07   &1.35           &621.45         &                 &         &          &               &               &                 \\
$f_{11}$ &1606.17   &                &622.60      &1607.72   &               &622.00         &622.60           &         &          &               &               &                 \\
         &1601.64   &4.53            &624.36      &1601.64   &6.08           &624.36         &                 &         &          &               &               &                 \\
         &1558.87   &4.64            &641.49      &1558.75   &4.29           &641.54         &                 &         &          &               &               &                 \\
$f_{12}$ &1554.23   &                &643.41      &1554.46   &               &643.31         &643.41           &         &          &               &               &                 \\
         &1550.39   &3.84            &645.00      &1550.41   &4.05           &644.99         &                 &         &          &               &               &                 \\
         &1505.24   &4.23            &664.34      &1505.34   &8.54           &664.43         &                 &         &          &               &               &                 \\
$f_{13}$ &1501.01   &                &666.22      &1496.80   &               &668.09         &666.22           &         &          &               &               &                 \\
         &1496.89   &4.12            &668.05      &1487.63   &9.17           &672.21         &                 &         &          &               &               &                 \\
         &1458.12   &4.01            &685.81      &1458.17   &4.13           &685.79         &                 &         &          &               &               &                 \\
$f_{14}$ &1454.11   &                &687.71      &1454.04   &               &687.74         &687.71           &         &          &               &               &                 \\
         &1449.89   &4.22            &689.71      &1449.80   &4.24           &689.75         &                 &         &          &               &               &                 \\
         &1416.59   &4.00            &705.92      &1417.80   &7.46           &705.32         &                 &         &          &               &               &                 \\
$f_{15}$ &1412.59   &                &707.92      &1410.34   &               &709.05         &707.92           &         &          &               &               &                 \\
         &1407.87   &4.72            &710.29      &1405.32   &5.02           &711.58         &                 &         &          &               &               &                 \\
         &1375.43   &4.63            &727.05      &1375.36   &4.58           &727.09         &                 &         &          &               &               &                 \\
$f_{16}$ &1370.80   &                &729.50      &1370.78   &               &729.51         &729.50           &         &          &               &               &                 \\
         &1367.11   &3.69            &731.47      &1367.15   &3.63           &731.45         &                 &         &          &               &               &                 \\
         &1332.55   &4.75            &750.44      &1332.34   &4.21           &750.56         &                 &         &          &               &               &                 \\
$f_{17}$ &1327.80   &                &753.12      &1328.13   &               &752.94         &753.12           &         &          &               &               &                 \\
         &1323.83   &3.97            &755.38      &1323.96   &4.17           &755.31         &                 &         &          &               &               &                 \\
         &1296.43   &4.05            &771.35      &          &               &               &                 &         &          &               &               &                 \\
$f_{18}$ &1292.38   &                &773.77      &1292.42   &               &773.74         &773.77           &         &          &               &               &                 \\
         &1287.58   &4.80            &776.65      &1287.55   &4.87           &755.31         &                 &         &          &               &               &                 \\
         &1267.99   &3.68            &788.65      &1265.41   &2.46           &790.26         &                 &         &          &               &               &                 \\
$f_{19}$ &1264.31   &                &790.94      &1262.95   &               &791.80         &790.94           &         &          &               &               &                 \\
         &1259.52   &4.79            &793.95      &1260.49   &2.46           &793.34         &                 &         &          &               &               &                 \\
         &1227.94   &4.13            &814.37      &1230.66   &3.05           &812.57         &                 &         &          &               &               &                 \\
$f_{20}$ &1223.81   &                &817.12      &1227.61   &               &814.58         &817.12           &         &          &               &               &                 \\
         &1219.54   &4.27            &819.98      &1223.39   &4.22           &817.40         &                 &         &          &               &               &                 \\
         &1194.73   &4.28            &837.01      &1197.12   &4.68           &835.34         &                 &         &          &               &               &                 \\
$f_{21}$ &1190.45   &                &840.02      &1192.44   &               &838.62         &840.02           &         &          &               &               &                 \\
         &1186.43   &4.02            &842.86      &1186.41   &6.03           &842.88         &                 &         &          &               &               &                 \\
\hline
\end{tabular}
\end{center}
\end{table*}

\begin{table*}
\begin{center}
\caption{The mode selections from Winget et al. (1991) and Costa et al. (2008) for $l$ = 2}
\begin{tabular}{lccccccccccccccc}
\hline
         &\multicolumn{3}{c}{Winget et al. (1991)}&\multicolumn{3}{c}{Costa et al. (2008)}   & Selected Period &         &\multicolumn{3}{c}{Costa et al. (2008)}   & Selected Period \\
\hline
ID       & Freq.    &$\delta\sigma$  & Per.       & Freq.    &$\delta\sigma$ & Per.          &$P_{\rm obs}$    &ID       & Freq.    &$\delta\sigma$ & Per.          &$P_{\rm obs}$    \\
$l$ = 2  &($\mu$Hz) &($\mu$Hz)       &(s)         &($\mu$Hz) &($\mu$Hz)      &(s)            &(s)              &         &($\mu$Hz) &($\mu$Hz)      &(s)            &(s)              \\
\hline
         &2961.15   &7.27            &337.71      &          &               &               &                 &         &1837.19   &5.86           &544.31         &                 \\
         &2953.88   &6.08            &338.54      &          &               &               &                 &         &1831.33   &3.18           &546.05         &                 \\
$f_{26}$ &2947.80   &                &339.24      &          &               &               &339.24           &$f_{38}$ &1828.15   &               &547.00         &                 \\
         &2940.26   &7.54            &340.11      &          &               &               &                 &         &1816.46   &11.69          &550.52         &                 \\
         &2933.28   &6.98            &340.91      &          &               &               &                 &         &          &               &               &                 \\
         &2863.00   &6.75            &349.28      &          &               &               &                 &         &1796.49   &7.52           &556.64         &                 \\
         &2856.25   &7.36            &350.11      &          &               &               &                 &         &1788.97   &9.58           &558.98         &                 \\
$f_{27}$ &2848.89   &                &351.01      &2851.03   &               &350.75         &351.01           &$f_{39}$ &1779.39   &               &561.99         &                 \\
         &2842.14   &6.75            &351.85      &2837.04   &13.99          &352.48         &                 &         &1774.68   &4.71           &563.48         &                 \\
         &2834.96   &7.18            &352.74      &2829.73   &7.31           &353.39         &                 &         &          &               &               &                 \\
         &          &                &            &          &               &               &                 &         &          &               &               &                 \\
         &2771.78   &7.50            &360.78      &2760.91   &9.05           &362.20         &                 &         &1750.73   &7.63           &571.19         &                 \\
$f_{28}$ &2764.28   &                &361.76      &2751.86   &               &363.39         &361.76           &$f_{40}$ &1743.10   &               &573.69         &                 \\
         &2757.34   &6.94            &362.67      &          &               &               &                 &         &1736.05   &7.05           &576.02         &                 \\
         &2750.92   &6.42            &363.51      &          &               &               &                 &         &1726.79   &9.26           &579.11         &                 \\
         &2589.72   &12.89           &386.14      &          &               &               &                 &         &1449.76   &7.36           &689.77         &                 \\
         &          &                &            &2584.45   &3.61           &386.93         &                 &         &1442.40   &7.33           &693.29         &                 \\
$f_{29}$ &2576.83   &                &388.07      &2580.84   &               &387.47         &388.07           &$f_{41}$ &1435.07   &               &696.83         &                 \\
         &2568.93   &7.90            &389.27      &          &               &               &                 &         &          &               &               &                 \\
         &2562.11   &6.82            &390.30      &2562.13   &11.29          &390.30         &                 &         &1416.83   &18.24          &705.80         &                 \\
         &          &                &            &2517.43   &10.60          &397.23         &                 &         &          &               &               &                 \\
         &          &                &            &2506.83   &7.20           &398.91         &                 &         &          &               &               &                 \\
$f_{30}$ &          &                &            &2499.63   &               &400.06         &                 &         &          &               &               &                 \\
         &          &                &            &          &               &               &                 &         &          &               &               &                 \\
         &          &                &            &2485.34   &14.29          &402.36         &                 &         &          &               &               &                 \\
         &          &                &            &2436.47   &9.29           &410.43         &                 &         &          &               &               &                 \\
         &2426.51   &6.97            &412.11      &2427.18   &6.69           &412.00         &                 &         &          &               &               &                 \\
$f_{31}$ &2419.54   &                &413.30      &2420.49   &               &413.14         &413.30           &         &          &               &               &                 \\
         &2413.12   &6.42            &414.40      &2413.30   &7.19           &414.37         &                 &         &          &               &               &                 \\
         &2406.18   &6.94            &415.60      &2406.22   &7.08           &415.59         &                 &         &          &               &               &                 \\
         &2366.76   &7.08            &422.52      &2366.58   &7.03           &422.55         &                 &         &          &               &               &                 \\
         &2359.68   &6.92            &423.79      &2359.55   &6.83           &423.81         &                 &         &          &               &               &                 \\
$f_{32}$ &2352.76   &                &425.03      &2352.72   &               &425.04         &425.03           &         &          &               &               &                 \\
         &2345.78   &6.98            &426.30      &2345.82   &6.90           &426.29         &                 &         &          &               &               &                 \\
         &2338.89   &6.89            &427.55      &2339.02   &6.80           &427.53         &                 &         &          &               &               &                 \\
         &          &                &            &2299.06   &8.42           &434.96         &                 &         &          &               &               &                 \\
         &2290.45   &7.35            &436.60      &2290.64   &               &436.56         &                 &         &          &               &               &                 \\
$f_{33}$ &2283.10   &                &438.00      &          &               &               &438.00           &         &          &               &               &                 \\
         &2276.69   &6.41            &439.23      &2276.61   &               &439.25         &                 &         &          &               &               &                 \\
         &2269.14   &7.55            &440.70      &2269.32   &7.29           &440.66         &                 &         &          &               &               &                 \\
         &          &                &            &2239.54   &6.85           &446.52         &                 &         &          &               &               &                 \\
         &          &                &            &2232.69   &7.65           &447.89         &                 &         &          &               &               &                 \\
$f_{34}$ &          &                &            &2225.04   &               &449.43         &449.43           &         &          &               &               &                 \\
         &          &                &            &2212.24   &12.80          &452.03         &                 &         &          &               &               &                 \\
         &          &                &            &2206.24   &6.00           &453.26         &                 &         &          &               &               &                 \\
         &          &                &            &1970.13   &9.58           &507.58         &                 &         &          &               &               &                 \\
         &          &                &            &1960.55   &7.35           &510.06         &                 &         &          &               &               &                 \\
$f_{35}$ &          &                &            &1953.20   &               &511.98         &                 &         &          &               &               &                 \\
         &          &                &            &1945.30   &7.90           &514.06         &                 &         &          &               &               &                 \\
         &          &                &            &1937.87   &7.43           &516.03         &                 &         &          &               &               &                 \\
         &1747.65   &8.91            &572.20      &          &               &               &                 &         &          &               &               &                 \\
         &1738.74   &6.15            &575.13      &          &               &               &                 &         &          &               &               &                 \\
$f_{36}$ &1732.59   &                &577.17      &          &               &               &                 &         &          &               &               &                 \\
         &1726.59   &6.00            &579.18      &          &               &               &                 &         &          &               &               &                 \\
         &1719.08   &7.51            &581.71      &          &               &               &                 &         &          &               &               &                 \\
         &1032.28   &6.05            &968.73      &1034.18   &16.55          &966.95         &                 &         &          &               &               &                 \\
         &1026.23   &8.13            &974.44      &          &               &               &                 &         &          &               &               &                 \\
$f_{37}$ &1018.10   &                &982.22      &1017.63   &               &982.68         &982.22           &         &          &               &               &                 \\
         &1011.81   &6.29            &988.33      &1011.43   &6.20           &988.70         &                 &         &          &               &               &                 \\
         &1004.77   &7.04            &995.25      &          &               &               &                 &         &          &               &               &                 \\
\hline
\end{tabular}
\end{center}
\end{table*}

\subsection{The model fittings on PG 1159-035}

\begin{table}
\begin{center}
\caption{The models with smallest values of $\sigma_{RMS}$. The ID numbers are consistent with that in Table 1.}
\begin{tabular}{lccccc}
\hline
ID   &$T_{\rm eff}$ &$M_{*}$      &log($M_{\rm env}/M_{*}$) &$\sigma_{RMS}$\\
     &(K)           &($M_{\odot}$)&                         &(s)           \\
\hline
1    &128600        &0.63         &-5.0                     &2.07          \\
2    &128200        &0.63         &-5.0                     &2.08          \\
3    &129000        &0.63         &-5.0                     &1.97          \\
4    &127400        &0.63         &-5.0                     &2.39          \\
5    &129400        &0.63         &-5.0                     &2.45          \\
\hline
\end{tabular}
\end{center}
\end{table}

\begin{figure}
\begin{center}
\includegraphics[width=9.0cm,angle=0]{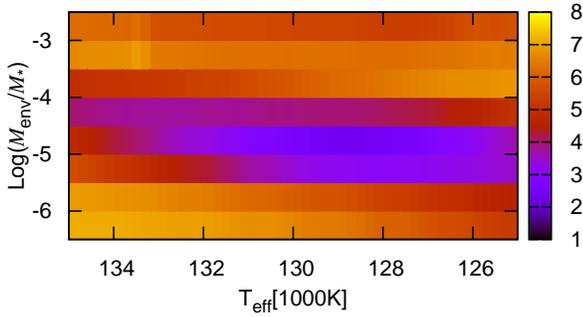}
\end{center}
\caption{The color residual diagram to fit PG 1159-035 with DOV star models. The value of $M_{\rm *}$ is 0.63\,$M_{\odot}$ for the models in the figure. The root-mean-square residual ($\sigma_{RMS}$) is 1.97\,s for the optimal model.}
\end{figure}

For each model, the eigenfrequencies are calculated and used to fit the 20 $l$ = 1, $m$ = 0 modes and 9 $l$ = 2, $m$ = 0 modes. An equation of root-mean-square residual ($\sigma_{RMS}$) is used to evaluate the fitting results, which is calculated by
\begin{equation}
\sigma_{RMS}=\sqrt{\frac{1}{n} \sum_{n}(P_{\rm obs}-P_{\rm cal})^2}.
\end{equation}
\noindent In Eq.\,(1), $n$ is the number of observed modes. It is 29 for PG 1159-035. The parameters of optimal models corresponding to each core composition (ID 1, 2, 3, 4, and 5) are displayed in Table 5. $T_{eff}$ is from 127400\,K to 129400\,K, $M_{*}$ is 0.630\,$M_{\odot}$, log($M_{\rm env}/M_{*}$) is -5.0, and $\sigma_{RMS}$ is from 1.97\,s to 2.45\,s. The difference of $T_{eff}$ in Table 5 is less than 1.6\%. However, the value of $\sigma_{RMS}$ can be improved by 20\%. Different core compositions (ID 1-5) have significant effects on the values of $\sigma_{RMS}$. We choose the evolving scenario of ID 3 as the optimal core composition. In Fig. 2, we show the color residual diagram to fit the 29 observed modes. The abscissa is the effective temperature, the ordinate is log($M_{\rm env}/M_{*}$). The colors show the fitting results. The model of $T_{eff}$ = 129000\,K (ID 3) and log($M_{\rm env}/M_{*}$) = -5.0 has the smallest $\sigma_{RMS}$ of 1.97\,s.

In Table 6, we show a comparison of the detailed fitting results for model 3 in Table 5. The 20 observed $l$ = 1 modes are represented by $P_{\rm obs}$($l$=1) and the 9 observed $l$ = 2 modes are represented by $P_{\rm obs}$($l$=2). The calculated modes are represented by $P_{\rm cal}$$(l,k)$. The value of $\sigma$ is $P_{\rm obs}$ minus $P_{\rm cal}(l,k)$ in seconds.

\begin{table*}
\begin{center}
\caption{The detailed fitting results with the model of ID 3 in Table 5. The value of $\sigma$ is $P_{\rm obs}$ minus $P_{\rm cal}(l,k)$ in seconds.}
\begin{tabular}{lllllllllllllllll}
\hline
$P_{\rm obs}$($l$=1)  &$P_{\rm cal}$$(l,k)$    &$\sigma$     &$P_{\rm obs}$($l$=2)  &$P_{\rm cal}$$(l,k)$ &$\sigma$    &$P_{\rm obs}$($l$=2)  &$P_{\rm cal}$$(l,k)$  &$\sigma$  \\
\hline
(s)                   &(s)                     &(s)          &(s)                   &(s)                  &(s)         &(s)                   &(s)                   &(s)       \\
\hline
               &351.48(1,15)            &             &               &329.19(2,25)         &            &               &738.31(2,56)          &          \\
               &373.15(1,16)            &             &339.24         &339.32(2,26)         &$\,$-0.08   &               &752.45(2,57)          &          \\
               &394.60(1,17)            &             &351.01         &350.43(2,27)         &$\,$ 0.58   &               &765.98(2,58)          &          \\
               &412.30(1,18)            &             &361.76         &364.07(2,28)         &$\,$-2.31   &               &778.72(2,59)          &          \\
430.04         &429.90(1,19)            &$\,$ 0.14    &               &377.26(2,29)         &            &               &792.55(2,60)          &          \\
450.71         &452.10(1,20)            &$\,$-1.39    &388.07         &389.62(2,30)         &$\,$-1.55   &               &807.66(2,61)          &          \\
469.57         &474.65(1,21)            &$\,$-5.08    &               &402.98(2,31)         &            &               &824.10(2,62)          &          \\
494.85         &495.06(1,22)            &$\,$-0.21    &413.30         &415.69(2,32)         &$\,$-2.39   &               &840.38(2,63)          &          \\
517.18         &518.15(1,23)            &$\,$-0.97    &425.03         &426.78(2,33)         &$\,$-1.75   &               &856.15(2,64)          &          \\
538.16         &540.48(1,24)            &$\,$-2.32    &438.00         &438.21(2,34)         &$\,$-0.21   &               &871.54(2,65)          &          \\
558.44         &559.48(1,25)            &$\,$-1.04    &449.43         &451.19(2,35)         &$\,$-1.76   &               &884.60(2,66)          &          \\
               &578.05(1,26)            &             &               &464.45(2,36)         &            &               &897.24(2,67)          &          \\
603.04         &598.31(1,27)            &$\,$ 4.73    &               &477.41(2,37)         &            &               &912.81(2,68)          &          \\
622.60         &620.41(1,28)            &$\,$ 2.19    &               &491.62(2,38)         &            &               &930.05(2,69)          &          \\
643.41         &641.18(1,29)            &$\,$ 2.28    &               &505.80(2,39)         &            &               &946.55(2,70)          &          \\
666.22         &664.51(1,30)            &$\,$ 2.23    &               &518.93(2,40)         &            &               &963.87(2,71)          &          \\
687.71         &688.16(1,31)            &$\,$-0.45    &               &531.90(2,41)         &            &982.22         &981.62(2,72)          &$\,$ 0.60 \\
707.92         &708.97(1,32)            &$\,$-1.05    &               &543.73(2,42)         &            &               &996.62(2,73)          &          \\
729.50         &729.78(1,33)            &$\,$-0.28    &               &555.69(2,43)         &            &               &                      &          \\
753.12         &750.50(1,34)            &$\,$ 2.62    &               &568.48(2,44)         &            &               &                      &          \\
773.77         &770.63(1,35)            &$\,$ 3.14    &               &582.90(2,45)         &            &               &                      &          \\
790.94         &791.88(1,36)            &$\,$-0.94    &               &597.58(2,46)         &            &               &                      &          \\
817.12         &816.53(1,37)            &$\,$ 0.59    &               &611.46(2,47)         &            &               &                      &          \\
840.02         &840.94(1,38)            &$\,$-0.92    &               &626.26(2,48)         &            &               &                      &          \\
861.72         &863.15(1,39)            &$\,$-1.43    &               &640.37(2,49)         &            &               &                      &          \\
               &887.71(1,40)            &             &               &651.87(2,50)         &            &               &                      &          \\
               &912.10(1,41)            &             &               &663.53(2,51)         &            &               &                      &          \\
               &932.21(1,42)            &             &               &677.91(2,52)         &            &               &                      &          \\
               &951.69(1,43)            &             &               &692.51(2,53)         &            &               &                      &          \\
               &974.05(1,44)            &             &               &707.35(2,54)         &            &               &                      &          \\
               &997.80(1,45)            &             &               &723.21(2,55)         &            &               &                      &          \\
\hline
\end{tabular}
\end{center}
\end{table*}

\begin{figure}
\begin{center}
\includegraphics[width=9.0cm,angle=0]{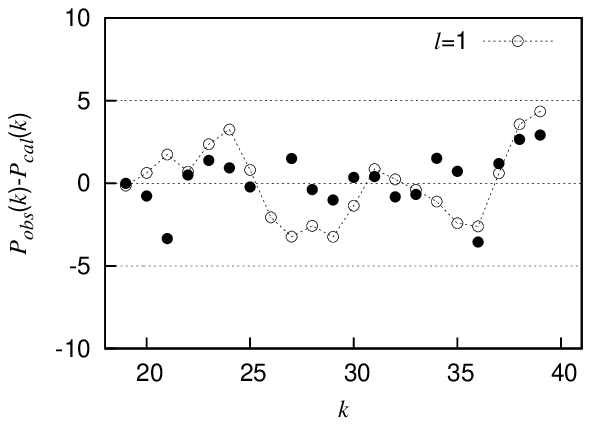}
\end{center}
\caption{Fitting the observed $l$ = 1 modes with the calculated modes in Table 6. The observed $l$ = 1 modes are represented by filled circles. The figure is drawn by subtracting a linear fitting function to the calculated modes from both the observed and calculated $l$ = 1 modes.}
\end{figure}

\begin{figure}
\begin{center}
\includegraphics[width=9.0cm,angle=0]{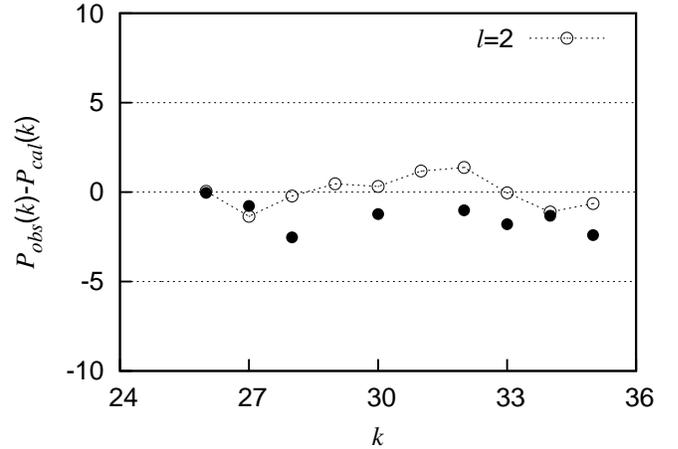}
\end{center}
\caption{Fitting the observed $l$ = 2 modes with the calculated modes in Table 6. The observed $l$ = 2 modes are represented by filled circles. The figure is drawn by subtracting a linear fitting function to the calculated modes from both the observed and calculated $l$ = 2 modes.}
\end{figure}

In Fig. 3, we show the fitting results for $l$ = 1 modes according to the modes in Table 6. Based on the optimal fitting model, the radial order $k$ is from 19 to 39 for the 20 observed $l$ = 1, $m$ = 0 modes except the mode of $k$ = 26. A linear function is fitted to the calculated modes of $k$ = 19-39. It is $P_{fit}(k)$ = 21.4384$\times$$k$ + 22.7058. By subtracting the linear function from both the observed and calculated modes, we made Fig. 3. The observed modes are represented by filled circles. The 20 observed $l$ = 1 modes are distributed near the horizontal zero line. The average period spacing for calculated $l$ = 1 modes is 21.44\,s, which is consistent with the average period spacing (21.5\,s) from Winget et al. (1991) and (21.43\,s) Costa et al. (2008). In Fig. 4 (a fitting function of $P_{fit}(k)$ = 12.5062$\times$$k$ + 14.1170), we show the fitting results for $l$ = 2 modes according to the modes in Table 6. The radial order $k$ is from 26 to 35 except 29 and 31. The average period spacing for calculated $l$ = 2 modes is 12.51\,s. It is consistent with the average period spacing (12.5\,s) from Winget et al. (1991) and (12.38\,s) Costa et al. (2008).

\begin{table}
\begin{center}
\caption{Spectral and asteroseismological results for PG 1159-035. The result of ID 1 is from Barstow et al. (1986) and Barstow \& Holberg (1990). The result of ID 2 is from Werner et al. (1991), Jahn et al. (2007), and Werner et al. (2011). The result of ID 3 is from Kawaler \& Bradley (1994). The result of ID 4 is from C\'orsico et al. (2008). The result of ID 5 is from this work.}
\begin{tabular}{lccccc}
\hline
ID   &$T_{\rm eff}$               &$M_{*}$                             &log$g$                   &$X_{C}/X_{He}/X_{O}$    \\
     &(K)                         &($M_{\odot}$)                       &                         &                        \\
\hline
1    &123000-124000               &                                    &                         &                        \\
2    &140000$\pm$5000             &                                    &\,7.0$\pm$0.5            &50/33/17                \\
3    &136000                      &0.59$\pm$0.01                       &\,7.38$\pm$0.1           &58/27/15                \\
4    &$128000^{+8600}_{-2600}$    &$0.565^{+0.025}_{-0.009}$           &$7.42^{+0.21}_{-0.12}$   &34.2/42.2/23.6          \\
5    &129000                      &0.63                                &\,7.59                   &50/33/17                \\
\hline
\end{tabular}
\end{center}
\end{table}

In Table 7, we show a comparison to previous spectral and asteroseismological work. With the X-ray spectrum from the EXOSAT observations, Barstow \& Holberg (1990) derived an effective temperature of 123000-124000\,K for PG 1159-035. With high-resolution ultraviolet spectra, Jahn et al. (2007) reported that PG 1159-035 had $T_{\rm eff}$ = 140000$\pm$5000 and log$g$ = 7.0$\pm$0.5. The main atmospheric constituent was $X_{C}/X_{He}/X_{O}$ = 50/33/17 (Werner et al. 1991, Werner et al. 2011).

With seed models from post-AGB stars, Kawaler \& Bradley (1994) evolved DOV star models by \texttt{WDEC}. Fitting the observed modes from Winget et al. (1991), they did asteroseismological study for PG 1159-035. They obtained an optimal model with $T_{\rm eff}$ = 136000\,K, $M_{*}$ = 0.59$\pm$0.01\,$M_{\odot}$, log$g$ = 7.38$\pm$0.1, and $X_{He}$ $\thickapprox$ 0.27 at the surface. Their models have assumed core compositions of 57\%C+43\%O, 40\%C+60\%O, 20\%C+80\%O, and so on. Fitting 20 observed $l$ = 1 modes,  Kawaler \& Bradley (1994) obtained a fitting error of $\sigma_{RMS}$ = 1.58\,s in Table 3 of their paper.

With the LPCODE evolutionary code (Althaus et al. 2005), C\'orsico et al. (2008) evolved MS stars (1.00-3.75\,$M_{\odot}$) to the post-AGB phase, through the very late thermal pulse and the resulting born-again episode, then to the PG 1159 stars. Fitting the observed modes identified by Winget et al. (1991) and Costa et al. (2008), C\'orsico et al. (2008) derived their optimal model with $T_{\rm eff}$ = $128000^{+8600}_{-2600}$\,K, $M_{*}$ = $0.565^{+0.025}_{-0.009}$\,$M_{\odot}$, log$g$ = $7.42^{+0.21}_{-0.12}$, and $X_{C}/X_{He}/X_{O}$ = 34.2/42.2/23.6.

We evolved grids of MS stars to white dwarfs at highest $T_{\rm eff}$ by \texttt{MESA}. The evolved core compositions are inserted into \texttt{WDEC} to evolve hot DOV star models. The atmospheric constituent was fixed to be the spectral value of $X_{C}/X_{He}/X_{O}$ = 50/33/17. According to the frequency splitting values for $l$ = 1 and 2 modes, 20 $l$ = 1 modes and 9 $l$ = 2 modes were selected based on the mode identifications of Winget et al. (1991) and Costa et al. (2008). Fitting the selected 29 observed modes, we obtained an optimal model with $T_{\rm eff}$ = 129000\,K, $M_{*}$ = 0.63\,$M_{\odot}$, log$g$ = 7.59, and $\sigma_{RMS}$ = 1.97\,s.

The atmospheric constituent $X_{C}/X_{He}/X_{O}$ = 50/33/17 are exactly consistent with the spectral results. The value of $T_{\rm eff}$ = 129000\,K and log$g$ = 7.59 are consistent with that of C\'orsico et al. (2008). The total stellar mass of $M_{*}$ = 0.63\,$M_{\odot}$ is a little larger than previous work. Kawaler \& Bradley (1994) and C\'orsico et al. (2008) identified the radial order of $f_{6}$ as 22. Costa et al. (2008) identified the radial order of $f_{6}$ as 20$\pm$2. According to our optimal model in Table 6, the radial order of $f_{6}$ is 23, being larger than previous work. Kawaler \& Bradley (1994) reported that log($M_{\rm env}/M_{*}$) was -2.40 for PG 1159-035. C\'orsico et al. (2008) reported that log($M_{\rm env}/M_{*}$) was -1.52. We obtained a very thin envelope of log($M_{\rm env}/M_{*}$) = -5.0, as shown in Fig. 2. This may be due to the use of fixed uniform envelopes with $X_{C}/X_{He}/X_{O}$ = 50/33/17. It is a preliminary study to evolve hot DOV star models by \texttt{WDEC} with core compositions from white dwarfs at highest $T_{\rm eff}$ evolved by \texttt{MESA}.

\subsection{The mode trapping effect and the rate of period change}

Studying the DAV stars, Winget, Van Horn \& Hansen (1981) first proposed that the composition stratified zones modified the wave propagation characteristics. The trapped or partly trapped modes produced. Trapped or partly trapped modes had their own asymptotic period spacings and therefore had minimal period spacings on the whole period versus period spacing diagram (Brassard et al. 1992). In this subsection, we discuss the mode trapping effect on the DOV star PG 1159-035. In Fig. 5 (Fig. 6), we show a diagram of $l$ = 1 modes ($l$ = 2 modes) for period versus period spacing and kinetic energy in the C/O core. The diagram is $P(k)$ versus $P(k+1)$ - $P(k)$. The minimum period spacings for $P(k)$ are actually in common to $P(k)$ and $P(k+1)$. The observed period spacings (filled circles) and calculated period spacings (open circles) are from modes in Table 6. Most of the observed modes are from Winget et al. (1991). The period versus period spacing diagram for the observed modes is consistent with that (Fig. 10) in Winget et al. (1991).

According to Fig. 5 and Fig. 6, together with Fig. 10 in Winget et al. (1991), there are five $l$ = 1 modes ($f_{3}$, $f_{7}$, $f_{10}$, $f_{14}$, and $f_{18}$ in Table 3) and two $l$ = 2 modes ($f_{27}$ and $f_{31}$ in Table 4) with minimum period spacings. In Fig. 5, the filled circles are basically matched by the open ones, except the modes of $f_{3}$, $f_{4}$, $f_{16}$, and $f_{18}$. For the calculated modes in Fig. 5, there are three modes with minimum kinetic energy distributed in the C/O core. This indicates that the uniform C/He/O atmosphere constituent in Fig. 1 may need to be improved for PG 1159-035 in the future work. For $l$ = 2 modes, there are two modes with minimum kinetic energy distributed in the C/O core. The two modes are basically consistent with the observed minimum period spacings in Fig. 6.

\begin{figure}
\begin{center}
\includegraphics[width=9.0cm,angle=0]{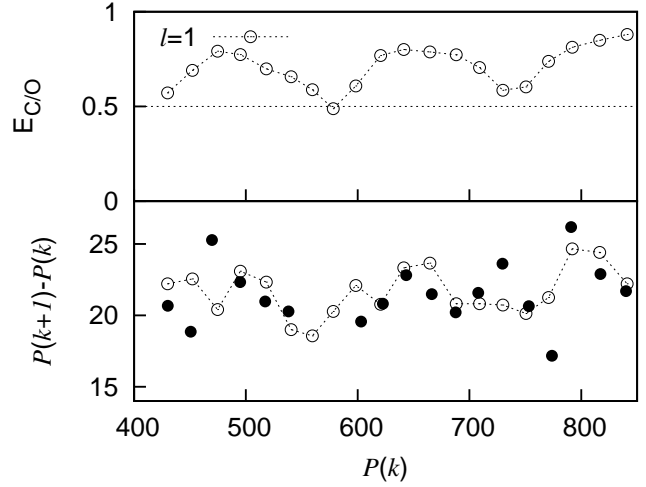}
\end{center}
\caption{Diagram of $l$ = 1 modes for period versus period spacing and kinetic energy in the C/O core. The calculated modes (open circles) are from model 3 in Table 5, while the observed period spacings are represented by filled circles.}
\end{figure}

\begin{figure}
\begin{center}
\includegraphics[width=9.0cm,angle=0]{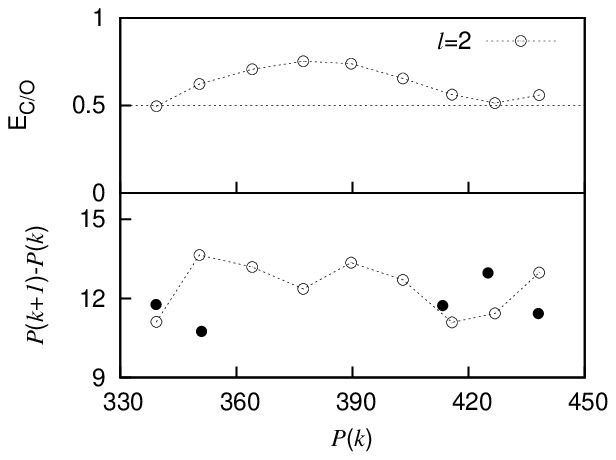}
\end{center}
\caption{Same as Fig. 5, but for $l$ = 2 modes.}
\end{figure}

\begin{table}
\begin{center}
\caption{The rates of period change for 8 observed $m$ = 0 modes from Costa \& Kepler (2008) and corresponding calculated rates of period change for our optimal model.}
\begin{tabular}{lllllllllllllllll}
\hline
$P_{\rm obs}$  &$l$  &$m$  &$\dot{P}_{\rm direct}$ &$\dot{P}_{\rm O-C}$  &$\dot{P}_{\rm cal}$\\
\hline
(s)            &     &     &($10^{-11}ss^{-1}$)    &($10^{-11}ss^{-1}$)  &($10^{-11}ss^{-1}$)\\
\hline
390.3 & 1 & 0 &$\,$ 10.4$\pm$2.3 &  $\,$ 12.596$\pm 0.020$ &  0.42-0.43                      \\
400.0 & 2 & 0 &$\,$ -6.8$\pm$2.8 &  $\,$ -0.123$\pm 0.027$ &  0.76-0.93                      \\
452.4 & 1 & 0 &$\,$  2.6$\pm$0.5 &  $\,$  4.720$\pm 0.007$ &  1.37-1.47                      \\
494.8 & 1 & 0 &$\,$-30.1$\pm$2.8 &  $\,$-30.526$\pm 0.093$ &  0.42-0.94                      \\
517.1 & 1 & 0 &$\,$ 18.2$\pm$0.8 &  $\,$ 15.172$\pm 0.045$ &  1.33-1.55                      \\
538.1 & 1 & 0 &$\,$  1.0$\pm$0.7 &  $\,$  4.304$\pm 0.010$ &  0.66-0.68                      \\
558.4 & 1 & 0 &$\,$ -4.3$\pm$1.9 &  $\,$-10.946$\pm 0.031$ &  -0.51-(-0.05)                  \\
561.9 & 2 & 0 &$\,$-20.5$\pm$3.7 &  $\,$-19.230$\pm 0.044$ &  0.15-0.70                      \\
\hline
\end{tabular}
\end{center}
\end{table}

\begin{figure}
\begin{center}
\includegraphics[width=9.0cm,angle=0]{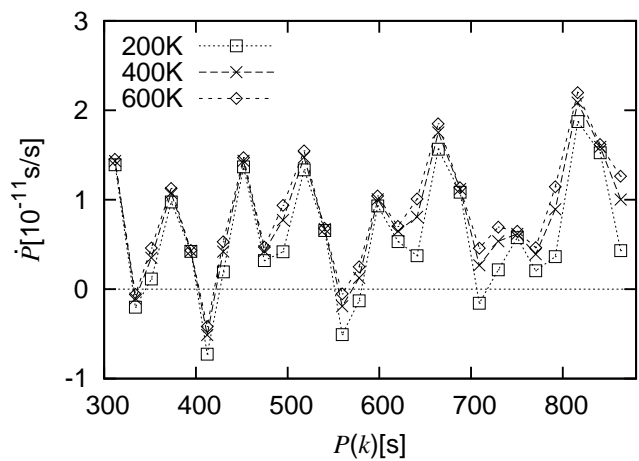}
\end{center}
\caption{Period versus rate of period change diagram for $l$ = 1 modes.}
\end{figure}

\begin{figure}
\begin{center}
\includegraphics[width=9.0cm,angle=0]{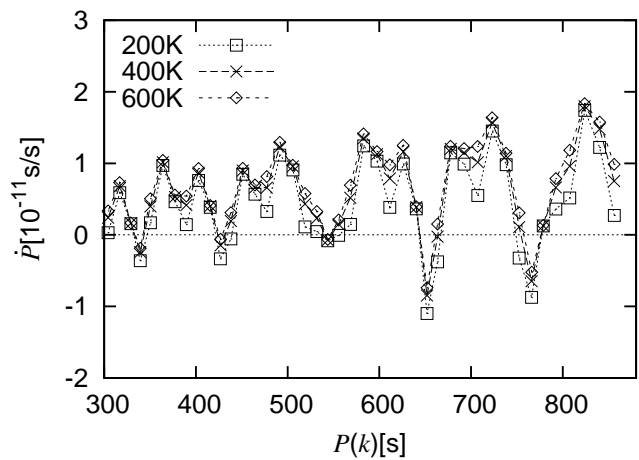}
\end{center}
\caption{Same as Fig. 7, but for $l$ = 2 modes.}
\end{figure}

The rate of period change ($\dot{P}$) for white dwarfs are related to the cooling of the star and the contraction of its envelope. The cooling of the star (maximum temperature) causes the increase of the pulsation periods. On the contrary, the contraction of the white dwarf envelope causes the decrease of the pulsation periods (Winget, Hansen \& Van horn 1983). Studying $\dot{P}$ for DAV stars G117-B15A and R548, Chen et al. (2017) obtained a coincidence result between the calculated $\dot{P}$ and observed $\dot{P}$ obtained by the O-C method. The DAV stars have $\dot{P}$ of $\sim$10$^{-15}$\,s/s. For DOV stars, the contraction process of the atmosphere is not negligible. The $\dot{P}$ is in the order of $\sim$10$^{-11}$\,s/s for DOV stars (Kepler et al. 2000). Based on the analysis on $\dot{P}$ of the 517\,s mode ($m$ = 0) for PG 1159-035, Winget \& Kepler (2008) reported that the $\dot{P}$ was still controlled by the stellar cooling process. Costa \& Kepler (2008) calculated the $\dot{P}$ for PG 1159-035 directly from the observed data and from the O-C method. They obtained the values of $\dot{P}$ for 27 modes. In Table 8, we list the values of $\dot{P}$ for 8 $m$ = 0 modes from Costa \& Kepler (2008). The observed rates of period change are from -30.5$\times$$10^{-11}ss^{-1}$ to 18.2$\times$$10^{-11}ss^{-1}$, with 4 positive values and 4 negative values.

Fitting PG 1159-035, both Kawaler \& Bradley (1994) and C\'orsico et al. (2008) obtained only positive rates of period change. In Fig. 7 and 8, we show the period versus $\dot{P}$ diagram for calculated $l$ = 1 and 2 modes. The calculated modes are from models around our optimal model with $T_{\rm eff}$ intervals of 200\,K, 400\,K, and 600\,K. The positive rates of period change dominate the figures, while negative rates of period change also exist. The calculated rates of period change are in the range of -1.1$\times$$10^{-11}ss^{-1}$ to 2.2$\times$$10^{-11}ss^{-1}$ in Fig. 7 and 8. The absolute values of calculated rates of period change are an order of magnitude smaller than the absolute values of observed rates of period change. The $l$ = 1 modes of 412\,s, 559\,s, 708\,s and $l$ = 2 modes of 339\,s, 426\,s have minimum rates of period change, which are corresponding to the modes with nearly minimum kinetic energy distributed in the C/O core in Fig. 5 and 6. Those modes have maximum kinetic energy distributed in the envelope and the contraction effect of the white dwarf atmosphere is obvious. Therefore, they have the minimum rates of period change in Fig. 7 and 8.

In the last column of Table 8, we list the corresponding calculated rates of period change. The 4 positive values of $\dot{P}_{\rm direct}$ or $\dot{P}_{\rm O-C}$ are fitted by 4 positive values of $\dot{P}_{\rm cal}$. The negative value ($\dot{P}_{\rm direct}$ or $\dot{P}_{\rm O-C}$) for the mode of 558.4\,s is fitted by a negative value. Fitting the other three modes with negative $\dot{P}_{\rm direct}$ or $\dot{P}_{\rm O-C}$, the calculated values of $\dot{P}_{\rm cal}$ are 0.76-0.93$\times$$10^{-11}ss^{-1}$, 0.42-0.94$\times$$10^{-11}ss^{-1}$, and 0.15-0.70$\times$$10^{-11}ss^{-1}$ respectively. Actually, fitting the 4 negative $\dot{P}_{\rm direct}$ or $\dot{P}_{\rm O-C}$, the calculated values of $\dot{P}_{\rm cal}$ have large dispersions among the models with $T_{\rm eff}$ intervals of 200\,K, 400\,K, and 600\,K. This may be due to the fact that the core compositions are from the white dwarf models with highest $T_{\rm eff}$ evolved by \texttt{MESA}.

With the core compositions from white dwarf models at highest $T_{\rm eff}$ evolved by \texttt{MESA}, the DOV star models evolved by \texttt{WDEC} have positive and negative rates of period change. It is consistent with the results showed in Fig. 1 of C\'orsico et al. (2008).

\section{Discussion and conclusions}

With high-speed photometric observations in 1989, Winget et al. (1991) made detailed mode identifications on PG 1159-035. With all available WET data from 1983, 1985, 1989, 1993, and 2002, Costa et al. (2008) identified 198 pulsation modes for PG 1159-035. Both of them obtained the frequency splitting values of $\delta\sigma_{l=1}$ $\sim$ 4.2\,$\mu$Hz and $\delta\sigma_{l=2}$ $\sim$ 6.9\,$\mu$Hz. According to the frequency splitting values, we select 20 $l$ = 1, $m$ = 0 modes and 9 $l$ = 2, $m$ = 0 modes to constrain the models, as shown in Table 3 and Table 4.

A grid of main-sequence stars are evolved to be white dwarfs by \texttt{MESA}. The core ($X_{C}$ = 0.65 as the boundary) compositions of the white dwarf models at highest $T_{\rm eff}$ were took out and added into \texttt{WDEC} to evolve hot DOV star models. Based on an EOS table of nearly pure C from Fontaine et al. (1977) and an EOS table of 50\% C and 50\% O from data files of \texttt{MESA}, we create an EOS table of pure O. An updated OPAL Rosseland mean opacity table of pure O (Iglesias \& Rogers 1996) is added into WDEC.  The atmospheric compositions is fixed to the spectral values of $X_{C}/X_{He}/X_{O}$ = 50/33/17. Then, grids of DOV star models are evolved with $M_{\rm *}$, $T_{\rm eff}$, and log($M_{\rm env}/M_{*}$) as grid parameters. The eigenfrequencies are calculated and used to fit the observed 20 $l$ = 1 modes and 9 $l$ = 2 modes for PG 1159-035.

We obtain an optimal model of $T_{eff}$ = 129000\,K, $M_{*}$ = 0.63\,$M_{\odot}$, log$g$ = 7.59, log($M_{\rm env}/M_{*}$) = -5.0, and $\sigma_{RMS}$ = 1.97\,s. The values of $T_{eff}$ and log$g$ are consistent with the values of C\'orsico et al. (2008). We fix the atmospheric constituent to the spectral values of $X_{C}/X_{He}/X_{O}$ = 50/33/17. The stellar mass is larger than previous results and the envelope mass is thinner than previous results. With the core compositions from white dwarf models at the highest $T_{eff}$ evolved by \texttt{MESA}, the calculated rates of period change for the optimal model have positive and negative values. The modes of minimum rates of period change correspond to the modes with minimum kinetic energy distributed in the C/O core and maximum kinetic energy distributed in the envelope. It is reasonable. For the work of observations, Costa \& Kepler (2008) identified 4 positive $\dot{P}$ and 4 negative $\dot{P}$ for 8 $m$ = 0 modes. The observed 4 positive $\dot{P}$ are fitted by calculated 4 positive $\dot{P}_{\rm cal}$. The observed 1 negative $\dot{P}$ is fitted by calculated 1 negative $\dot{P}_{\rm cal}$. There were only positive rates of period change for previous asteroseismological work.

\section{Acknowledgements}

This work is supported by the National Natural Science Foundation of China (Grant Nos. 11803004, and 11563001). Thank the supporting by the Youth Talent Lifting Project of Yunnan Association for Science \& Technology. We are very grateful to J. Su, Q. S. Zhang, and Y. Li for their kindly discussion and suggestions.

\label{lastpage}

\end{document}